\documentclass[conference,compsoc]{IEEEtran}

\usepackage{cite}
\usepackage{amsmath}
\usepackage{amssymb}
\usepackage{amsfonts}
\usepackage{graphicx}
\usepackage{textcomp}
\usepackage[table]{xcolor}
\usepackage{url}
\usepackage{array}
\usepackage{booktabs}
\usepackage{longtable}
\usepackage{multirow}
\usepackage{tabularx}
\usepackage{makecell}
\usepackage{float}
\usepackage{enumitem}
\usepackage{balance}
\usepackage{hyperref}
\hypersetup{colorlinks=true, linkcolor=blue, citecolor=blue, urlcolor=blue}
\newcolumntype{L}[1]{>{\raggedright\arraybackslash}p{#1}}
\newcolumntype{C}[1]{>{\centering\arraybackslash}p{#1}}
\newcolumntype{Y}{>{\raggedright\arraybackslash}X}

\newcommand{\levelLow}{\cellcolor{green!12}Low}
\newcommand{\levelMed}{\cellcolor{yellow!18}Medium}
\newcommand{\levelHigh}{\cellcolor{orange!20}High}
\newcommand{\levelVHigh}{\cellcolor{orange!32}Very High}

\begin{document}

\title{SoK: Security of Autonomous LLM Agents in Agentic Commerce}

\author{
\IEEEauthorblockN{
Qian'ang Mao\IEEEauthorrefmark{1},
Jiaxin Wang\IEEEauthorrefmark{1},
Ya Liu\IEEEauthorrefmark{1},
Li Zhu\IEEEauthorrefmark{1},
Cong Ma\IEEEauthorrefmark{2}\IEEEauthorrefmark{3},
Jiaqi Yan\IEEEauthorrefmark{1}
}
\IEEEauthorblockA{\IEEEauthorrefmark{1}Nanjing University}
\IEEEauthorblockA{\IEEEauthorrefmark{2}Southern University of Science and Technology}
\IEEEauthorblockA{\IEEEauthorrefmark{3}City University of Hong Kong}
}

\maketitle

\begin{abstract}
Autonomous large language model (LLM) agents such as OpenClaw are pushing agentic commerce from human-supervised assistance toward machine actors that can negotiate, purchase services, manage digital assets, and execute transactions across on-chain and off-chain environments. Protocols such as the Trustless Agents standard (ERC-8004), Agent Payments Protocol (AP2), OKX Agent Payments Protocol (APP), the HTTP 402-based payment protocol (x402), Agent Commerce Protocol (ACP), the Agentic Commerce standard (ERC-8183), and Machine Payments Protocol (MPP) enable this transition, but they also create an attack surface that existing security frameworks do not capture well. This Systematization of Knowledge (SoK) develops a unified security framework for autonomous LLM agents in commerce and finance. We organize threats along five dimensions: agent integrity, transaction authorization, inter-agent trust, market manipulation, and regulatory compliance. From a systematically curated public corpus of academic papers, protocol documents, industry reports, and incident evidence, we derive 12 cross-layer attack vectors and show how failures propagate from reasoning and tooling layers into custody, settlement, market harm, and compliance exposure. We then propose a layered defense architecture addressing authorization gaps left by current agent-payment protocols. Overall, our analysis shows that securing agentic commerce is inherently a cross-layer problem that requires coordinated controls across LLM safety, protocol design, identity, market structure, and regulation. We conclude with a research roadmap and a benchmark agenda for secure autonomous commerce.
\end{abstract}


\section{Introduction}
\label{sec:intro}

The financial industry has long been at the forefront of adopting computational automation, from early algorithmic trading systems~\cite{greenwald2001Bidding, vetsikas2003Designing} to modern high-frequency trading platforms. However, a new paradigm is emerging that fundamentally alters the relationship between artificial intelligence (AI) and financial decision-making: \emph{fully autonomous large language model (LLM)-based agents} that operate without continuous human oversight, control their own digital wallets or payment credentials, and execute financial transactions independently~\cite{steinberger2025Openclaw, mony2025Llm, rizinski2026AiAgents}.

Unlike traditional automated trading systems that follow pre-programmed rules, these agents leverage the reasoning, planning, and natural language understanding capabilities of large language models to interpret market conditions, negotiate with counterparties (including other agents), and adapt their strategies in real time~\cite{ding2024Llm, nie2024Survey, lee2024Survey}. Projects such as OpenClaw~\cite{steinberger2025Openclaw} (formerly Clawdbot) exemplify this trend, providing open-source frameworks for deploying LLM agents that can autonomously manage cryptocurrency portfolios, execute decentralized finance (DeFi) trades, and interact with smart contracts on Ethereum and other blockchains~\cite{mony2025Llm, xiao2025TradingWith}.

This shift toward full autonomy is accelerated by the emergence of machine-to-machine payment protocols. Ethereum's Trustless Agents standard (ERC-8004) enables agents to hold and transfer tokens through standardized smart contract interfaces~\cite{goenka2026Tesspay}. The Agent Payments Protocol (AP2) provides a framework for authenticated, verifiable payments between autonomous agents~\cite{liu2026Agenticpay}. OKX's Agent Payments Protocol (APP) similarly targets agent-to-agent and agent-to-service commerce, extending payment flows with negotiation, metering, escrow, and dispute-resolution concepts~\cite{okx2026AgentPaymentsProtocol}. The HTTP 402-based payment protocol (x402) embeds payment capabilities directly into HTTP requests, enabling agents to pay for API calls, data feeds, and computational resources without human authorization~\cite{goenka2026Tesspay}. Tempo's deployment of the Machine Payments Protocol (MPP) extends this model with a rail-agnostic challenge--credential--receipt flow over HTTP 402, supporting both one-time charges and session-based pay-as-you-go channels for APIs, Model Context Protocol (MCP) tools, and streamed services~\cite{tempo2026MachinePayments, mpp2026Protocol}. Together, these protocols form the infrastructure of an emerging \emph{agentic economy}~\cite{birch2025Agentic, zhang2026Agentic} in which billions of dollars may flow through agent-mediated channels with minimal human oversight.


Despite the rapid growth of this ecosystem, security research remains fragmented across several disconnected communities. The LLM security community focuses on prompt injection, jailbreaking, and alignment~\cite{greshake2023Indirect}, but often treats financial applications as merely another use case without accounting for the unique properties of financial systems (irreversibility of transactions, regulatory requirements, systemic risk). The blockchain security community addresses smart contract vulnerabilities and DeFi exploits~\cite{acharya2025Secure} but has yet to grapple with the implications of LLM-controlled digital wallets and payment credentials. The financial technology (FinTech) research community examines AI-driven trading strategies and investment management applications~\cite{cao2025Survey, zhao2024Revolutionizing, kong2024Large} but largely assumes human oversight as a given. The multi-agent systems community has a rich history of studying agent-mediated commerce~\cite{guttman1998Agent, he2003AgentMediated, sierra2004Agent} but its frameworks predate the capabilities and vulnerabilities of LLM-based agents.


This fragmentation creates dangerous blind spots. An autonomous financial agent is simultaneously an LLM (vulnerable to prompt injection), a blockchain or payment-network actor (subject to settlement and execution risks), a financial intermediary (bound by regulatory requirements), and a participant in a multi-agent ecosystem (susceptible to strategic manipulation by other agents). No existing security framework addresses this full stack of concerns. A compromise at any layer, whether it is a prompt injection that triggers an unauthorized trade, a malicious integration that drains an agent's digital wallet, or a coordinated attack by adversarial agents that manipulates market prices, can have cascading consequences that propagate across layers~\cite{acharya2025Secure}.


This paper makes the following contributions:

\begin{itemize}[leftmargin=*]
\item \textbf{Unified Threat Taxonomy.} We present a comprehensive taxonomy of security threats specific to autonomous LLM agents in financial automation, organized across five dimensions: agent integrity, transaction authorization, inter-agent trust, market manipulation, and regulatory compliance (\S\ref{sec:framework}).

\item \textbf{Cross-Layer Analysis.} We analyze how vulnerabilities at one layer (e.g., prompt injection at the LLM layer) propagate into harm at another (e.g., unauthorized token transfers at the blockchain layer), and we characterize 12 cross-layer attack vectors together with their adversary preconditions and mitigations (\S\ref{sec:comparison}).

\item \textbf{Protocol Security Assessment.} We assess emerging agent-payment protocols (ERC-8004, AP2, OKX APP, x402, MPP) from the perspective of autonomous deployment, identifying protocol-level weaknesses that are manageable in human-operated settings but dangerous in autonomous ones (\S\ref{sec:framework}).

\item \textbf{Defense Framework.} We propose a compact layered defense architecture spanning prompt hardening, payment authorization, tool provenance, decentralized identity, and market-level safeguards, and we relate its coverage to the threat taxonomy (\S\ref{sec:trends}).

\item \textbf{Corpus-Grounded Synthesis.} We assemble and analyze a systematically curated public corpus spanning academic papers, protocol documents, industry reports, and incident evidence, and use it to ground the threat taxonomy, protocol assessment, and comparative analysis.
\end{itemize}

\section{Background and Terminology}
\label{sec:background}

\subsection{Defining Autonomous Financial Agents}

The term ``AI agent'' has been applied to systems ranging from simple chatbots to sophisticated multi-step planners~\cite{maes1994Agents, jarrahi2025Rethinking}. We adopt a precise definition tailored to the financial domain:

\begin{quote}
\textit{An autonomous financial agent is a software system powered by one or more large language models that (1) maintains persistent state including financial assets and payment instruments such as digital wallets, accounts, or delegated payment credentials, (2) independently plans and executes financial transactions, (3) operates without requiring per-transaction human approval, and (4) interacts with external systems including blockchains, payment networks, exchanges, and other agents.}
\end{quote}

This definition intentionally uses \emph{digital wallets} as an umbrella term that includes on-chain wallets, custodial stored-value accounts, and delegated payment credentials rather than only crypto-native custody.

This definition excludes AI-assisted trading tools that require human confirmation (which we term \emph{co-pilot systems}~\cite{adedoyin2025Human, kong2024Large}), traditional algorithmic trading bots that follow fixed rules (which we term \emph{programmatic traders}~\cite{greenwald2001Bidding}), and LLM-based chatbots that provide financial advice but cannot execute transactions (which we term \emph{advisory agents}~\cite{desai2023Llms, kong2024Large}).

\subsection{Key Systems and Frameworks}

\subsubsection{OpenClaw and Clawdbot}
OpenClaw~\cite{steinberger2025Openclaw} (formerly known as Clawdbot) is an open-source framework that enables the deployment of fully autonomous LLM agents with blockchain-based digital wallet capabilities. It provides a modular architecture where LLM reasoning is connected to blockchain transaction execution through a plugin system. OpenClaw exemplifies the ``agent-as-wallet-holder'' paradigm in which the LLM directly controls private keys or has delegated signing authority~\cite{rizinski2026AiAgents, acharya2025Secure}.

\subsubsection{ERC-8004}
ERC-8004 is an Ethereum standard that defines a smart contract interface for agent-controlled token operations~\cite{goenka2026Tesspay}. Unlike traditional ERC-20 token transfers that assume a human signer, ERC-8004 introduces agent identity verification, spending limits, and revocation mechanisms designed for machine-to-machine interactions. The standard enables smart contracts to distinguish between human-initiated and agent-initiated transactions and apply different authorization policies accordingly.

\subsubsection{Agent Payments Protocol (AP2)}
AP2~\cite{liu2026Agenticpay} is a protocol layer built on top of existing payment rails (both blockchain and traditional) that provides standardized mechanisms for agent-to-agent payment negotiation, execution, and settlement. AP2 introduces the concept of \emph{payment intents}, which are machine-readable descriptions of desired payment outcomes that agents can negotiate over before committing to a transaction.

\subsubsection{OKX Agent Payments Protocol (APP)}
OKX's Agent Payments Protocol (APP) is an open standard proposed by OKX Onchain OS for agent commerce across chains~\cite{okx2026AgentPaymentsProtocol, okx2026AgentPaymentsProtocolDocs}. APP is explicitly broader than single-request payment protocols: it describes agent communication and negotiation, service payments, agent-to-agent payments, top-up and deduct billing, plan-based payments, and escrow-mediated commerce. The protocol documentation defines four payment intents (\texttt{charge}, \texttt{escrow}, \texttt{session}, and \texttt{upto}) and a Broker role that stores state, mints payment identifiers, verifies buyer credentials against stored challenges, submits settlement transactions, and exposes status queries for both sides~\cite{okx2026AgentPaymentsProtocolDocs}. Because dispute handling and escrow are still presented inconsistently across launch materials and implementation documentation, APP is best treated as an emerging commerce protocol whose security properties require specification-level validation rather than only launch-announcement evidence.

\subsubsection{Virtuals Protocol and Agent Commerce Protocol (ACP-Commerce)}
\textbf{Terminology note.} We disambiguate three overloaded acronyms throughout this paper. \textbf{ACP} as used in this paper refers exclusively to the \emph{Agent Commerce Protocol} by Virtuals Protocol~\cite{virtuals2026ACP}, which is a settlement and escrow coordination layer (we call this \textbf{ACP-Commerce} when disambiguation is needed). Unrelated concurrent work uses ``ACP'' for an \emph{Agent Control Protocol}, which is a deterministic pre-action authorization fabric~\cite{acpcontrol2026Admission}, and we denote that protocol as \textbf{ACP-Control}. Similarly, \textbf{PDR} in this paper means \emph{Payment Delivery Receipt} (a post-settlement cryptographic proof of payment completion, as formalized in~\cite{onchain2026AgentSoK}); unrelated literature uses PDR for \emph{Policy Decision Record}. We use ``PDR'' exclusively in the payment-delivery sense throughout.

Virtuals Protocol~\cite{virtuals2024Whitepaper} is a decentralized infrastructure platform built on Base (Ethereum Layer 2) that enables the creation, co-ownership, tokenization, and monetization of autonomous AI agents. Its cognitive engine, the GAME (Generative Agents with Modular Execution) framework~\cite{virtuals2025GAME}, provides a modular decision-making architecture separating task planning from execution.

Central to Virtuals is the \emph{Agent Commerce Protocol} (ACP)~\cite{virtuals2026ACP}, a standardized coordination and settlement layer for agent-to-agent commerce. ACP operates in four phases: (1) \emph{negotiation}, where agents agree on terms and produce a cryptographically signed Proof of Agreement; (2) \emph{transaction}, where payments and deliverables are held in escrow; (3) \emph{evaluation}, where specialized evaluator agents assess whether deliverables meet terms; and (4) \emph{settlement}, where funds are released or returned based on evaluation. This protocol introduces a novel trust primitive, namely the \emph{evaluator agent}, that enables trust in subjective or non-deterministic tasks but simultaneously introduces a new attack surface if the evaluator itself is compromised.

\subsubsection{ERC-8183: Agentic Commerce Standard}
Building on ACP's operational experience, Crapis et al.\ proposed ERC-8183~\cite{virtuals2026ERC8183} in March 2026 as an Ethereum standard for trustless commercial transactions between AI agents. ERC-8183 defines a core ``Job'' primitive with three roles (Client, Provider, Evaluator) and a state machine (Open $\to$ Funded $\to$ Submitted $\to$ Terminal). The standard is extensible via hooks, which are optional smart contracts for custom logic such as milestone payments, bidding, and reputation updates, and integrates with ERC-8004 for portable on-chain reputation. The proposal was motivated by what the ERC-8183 authors describe as over \$3M in agent-to-agent transactions observed on the Virtuals/ACP platform without any escrow or verification mechanism~\cite{virtuals2026ERC8183}, a figure that is unaudited but directionally indicative of the scale of unprotected commerce.

\subsubsection{x402 Protocol}
The x402 protocol, initiated by Coinbase and analyzed in the agentic-commerce context by~\cite{goenka2026Tesspay}, embeds payment capabilities into the HTTP protocol itself. When an agent makes an HTTP request to a resource that requires payment, the server responds with a 402 Payment Required status code along with machine-readable payment instructions. The agent can then autonomously fulfill the payment and retry the request. This protocol enables seamless pay-per-use access to APIs, data feeds, and computational services without pre-established billing relationships.

\subsubsection{Tempo and Machine Payments Protocol (MPP)}
Tempo is a payments-first blockchain optimized for low-cost stablecoin settlement and inline machine payments~\cite{tempo2026MachinePayments}. On top of Tempo, the Machine Payments Protocol (MPP) is an open standard co-authored by Stripe and Tempo that standardizes a challenge--credential--receipt flow over HTTP 402 and extends it to MCP transports~\cite{weinstein2026IntroducingMachinePayments, mpp2026Protocol}. On Tempo, MPP supports both \texttt{charge} intents for one-time payments and \texttt{session} intents that open escrow-backed channels and use off-chain signed vouchers for near-zero-latency pay-as-you-go billing~\cite{tempo2026MachinePayments, mpp2026Protocol}. This design makes MPP especially relevant for monetized APIs, MCP tool invocations, and streamed AI services. Adjacent protocol proposals are already exploring privacy-preserving settlement and explicit human-override semantics for agent commerce, as illustrated by AESP~\cite{aesp2026HumanSovereign}. At the implementation layer, these payment flows also intersect with lower-level signing primitives such as typed structured-data signing for wallet- or credential-bound intents and signed HTTP request binding~\cite{eip712, rfc9421}.

These contemporary protocols extend a much older line of agent-mediated payment research, including secure delegated payment schemes for software and mobile agents~\cite{pang2002SecurePayment, wang2005MobilePayment}. What is new in the current setting is not autonomous payment itself, but the combination of autonomous payment with open-ended LLM reasoning, untrusted tool use, and natural-language attack surfaces.

\subsubsection{Model Context Protocol (MCP)}
MCP~\cite{research2024Model, security2025Model} is an open protocol that standardizes how LLM agents interact with external tools, data sources, and services. In the financial context, MCP serves as the primary interface through which agents access market data, execute trades, and invoke smart contract functions. MCP's security properties, or the lack thereof, directly impact the security of financial operations conducted through it.

\subsubsection{Protocol Deployment Status and Maturity}

We explicitly qualify the deployment status of the agent payment protocols analyzed in this SoK, as of early 2026, to distinguish deployed behavior from proposed features. ERC-8004 and ERC-8183 are Ethereum Improvement Proposals in draft/community review status with limited on-chain deployment. AP2 is a research proposal~\cite{liu2026Agenticpay} with no widely adopted reference implementation. OKX APP is a newly announced open-standard and SDK-oriented protocol stack whose escrow and dispute-resolution features are still described as forthcoming~\cite{okx2026AgentPaymentsProtocol}. x402 has early adopter deployment by Coinbase and a growing ecosystem of MCP-compatible payment middleware, but is not yet standardized. MPP is co-authored by Stripe and Tempo and has a live Tempo mainnet deployment with documented API support~\cite{tempo2026MachinePayments, mpp2026Protocol}; it is the most operationally mature of the group. ACP/ERC-8183 is deployed on Virtuals Protocol's platform but remains Virtuals-specific. Our security analysis throughout the paper covers both deployed behavior (where independently verifiable) and proposed features (where explicitly noted). Claims about security properties of deployed behavior are grounded in protocol specifications and public chain data; claims about proposed features are explicitly speculative.

Protocol maturity is discussed comparatively in \S\ref{sec:comparison}.

\subsection{Levels of Agent Autonomy}

Drawing on prior discussions of agent autonomy and principal-agent dynamics~\cite{stocker2025PrincipalAgent, kapoor2024AgentsThat}, we distinguish four levels of agent autonomy in financial operations:

\begin{itemize}[leftmargin=*]
\item \textbf{Level 0 (Advisory):} The agent analyzes data and provides recommendations; all actions are taken by humans~\cite{kong2024Large, desai2023Llms}.
\item \textbf{Level 1 (Supervised):} The agent can propose and execute pre-approved transaction types within strict limits; humans approve exceptions~\cite{adedoyin2025Human, ding2024Llm, xiao2025TradingWith}.
\item \textbf{Level 2 (Delegated):} The agent independently executes a broad range of transactions within policy constraints; humans review periodically~\cite{ding2024Llm}.
\item \textbf{Level 3 (Fully Autonomous):} The agent independently manages a portfolio or financial operation with no per-transaction human oversight; humans set high-level goals and constraints only~\cite{steinberger2025Openclaw, mony2025Llm}.
\end{itemize}

This paper primarily concerns the security challenges of Level 2 and Level 3 agents, as these levels introduce qualitatively new risks that do not exist in supervised settings.

\subsection{Scope and Boundaries}

Our analysis focuses on security threats that arise specifically from the \emph{intersection} of agent autonomy and financial operations. We deliberately exclude:

\begin{itemize}[leftmargin=*]
\item \textbf{Generic LLM vulnerabilities} (e.g., hallucination, bias) except where they have specific financial security implications.
\item \textbf{Traditional financial risks} (e.g., market risk, credit risk) except where agent autonomy fundamentally changes their character.
\item \textbf{Regulatory compliance in isolation} except where autonomous agent behavior creates novel compliance challenges.
\end{itemize}

Our primary focus is blockchain-based and API-based agentic commerce (MCP, x402, MPP, ACP), reflecting the current frontier of autonomous agent deployment with real-asset exposure. Traditional payment rails and cross-chain operations are noted as important but out-of-scope extensions and are summarized briefly in \S\ref{sec:gaps}.

\section{Methodology}
\label{sec:methodology}

\subsection{Literature Collection}

Our systematization draws on literature from five intersecting research communities: (1) LLM security and alignment, (2) autonomous agent architectures, (3) blockchain and DeFi security, (4) financial technology and algorithmic trading, and (5) multi-agent systems and mechanism design.

We searched Google Scholar and Web of Science using 23 phrases spanning agentic-commerce core terms, autonomous payments and payment protocols, delegation and authorization, Model Context Protocol security, prompt injection and agent security, Web3 custody, financial LLMs, autonomous trading, and historical agent-mediated e-commerce. Because protocol specifications, regulatory materials, industry reports, and implementation documents central to agentic-commerce security are unevenly indexed in scholarly databases, we supplemented the database search with backward snowballing and targeted inclusion of these non-traditional sources.

\subsection{Selection Criteria}

We included works that directly address autonomous financial agents, agent-payment or settlement protocols, attacks and defenses relevant to autonomous execution, empirical evidence of agent behavior in financial settings, or theoretical foundations for trust, identity, and authorization. We excluded papers on generic LLM capability without financial-security relevance, traditional algorithmic trading without agent autonomy, and non-technical policy discussion without concrete security content. We also retained foundational multi-agent commerce work where it directly informs today's agentic-commerce threat model~\cite{guttman1998Agent, guttman1999Agent, sandholm1999Automated, he2003AgentMediated}.

\subsection{Selection Process and Evidence Base}

We followed a PRISMA-style selection flow. The database stage yielded 1,373 records, which were reduced to 1,237 candidates after DOI- and title-level deduplication. We then screened titles, abstracts, and full texts against the inclusion and exclusion criteria. Database retrieval contributed 37 works to the current public corpus, 105 additional works were retained through backward snowballing and targeted inclusion of protocol documents, regulatory texts, industry reports, implementation notes, and other poorly indexed materials, and the remaining 1,192 database-originated candidates were screened out. The resulting public corpus forms the evidentiary basis for the analysis in this paper.

For the released 30-row blinded replication set, two independent coders assigned source and target layers. On the 17 rows where both coders provided non-empty source and target labels, agreement was $\kappa = 0.850$ for source-layer labels, $\kappa = 0.833$ for target-layer labels, and $\kappa = 0.871$ for the joint ordered $(\text{source}, \text{target})$ pair.

Table~\ref{tab:vector_evidence} lists the currently released per-vector supporting works in the public corpus, distinguishing direct instantiation (confirmed incident, PoC, or experimental demonstration) from derived support (used as conceptual precursor or cross-paper synthesis support in the released mapping).

\begin{table*}[t]
\centering
\scriptsize
\setlength{\tabcolsep}{3pt}
\renewcommand{\arraystretch}{1.03}
\caption{Per-vector released supporting works in the current public corpus.}
\label{tab:vector_evidence}
\begin{tabular}{L{0.75cm}L{2.25cm}L{6.0cm}L{6.0cm}}
\toprule
\textbf{Vector} & \textbf{Name} & \textbf{Direct} & \textbf{Derived} \\
\midrule
P2T & Prompt-to-Transaction & Greshake et al.~\cite{greshake2023Indirect}; Acharya~\cite{acharya2025Secure}; Nieper-Wisskirchen et al.~\cite{preprint2026Security} & No released derived-support evidence in the current snapshot. \\
T2R & Tool-to-Reasoning & Model Context Protocol~\cite{security2025Model}; Maloyan and Namiot~\cite{mcpanalysis2026MCP}; Zhang et al.~\cite{agentaudit2026Audit} & No released derived-support evidence in the current snapshot. \\
A2M & Agent-to-Market & Allouah et al.~\cite{allouah2025What}; Kapoor et al.~\cite{kapoor2025Build} & Liu et al.~\cite{liu2026Agenticpay}; Chung and Honavar~\cite{chung2000NegotiationModel}; de Paula et al.~\cite{de2001Bilateral}; Cai et al.~\cite{guo2025Findebate} \\
T2T & Tool-to-Transaction & Acharya~\cite{acharya2025Secure}; Shittu~\cite{shittu2025VirtualsProtocolBug}; Ruan et al.~\cite{ruan2025PracticalExploit} & Deng et al.~\cite{openclaw2026Lifecycle} \\
P2K & Prompt-to-Key & Acharya~\cite{acharya2025Secure} & Steinberger~\cite{steinberger2025Openclaw}; Luo et al.~\cite{mony2025Llm}; Rizinski and Trajanov~\cite{rizinski2026AiAgents} \\
C2E & Collusion-to-Escrow & Virtuals Protocol~\cite{virtuals2026ACP}; Crapis et al.~\cite{virtuals2026ERC8183} & Liu et al.~\cite{liu2026Agenticpay}; Yu et al.~\cite{he2025TrustAgent}; de Witt~\cite{multiagent2025Security}; Hu and Rong~\cite{interagenttrust2025Taxonomy} \\
O2P & Oracle-to-Position & Moreno~\cite{moreno2025Predicting}; Assis et al.~\cite{assis2024Analysis} & Nabar and Shroff~\cite{nabar2023Conservative}; Kim et al.~\cite{kim2023Gans} \\
N2C & Neg-to-Compliance & Faysal et al.~\cite{faysal2026Agentic} & Liu et al.~\cite{liu2026Agenticpay}; Allouah et al.~\cite{allouah2025What}; Hornuf et al.~\cite{hornuf2025Making} \\
I2M & Identity-to-Market & Xu et al.~\cite{ping2005Deceit} & No released derived-support evidence in the current snapshot. \\
S2I & Supply-to-Integrity & Model Context Protocol~\cite{security2025Model}; Ruan et al.~\cite{ruan2025PracticalExploit} & Maloyan and Namiot~\cite{mcpanalysis2026MCP}; Zhang et al.~\cite{agentaudit2026Audit} \\
M2A & Model-to-Authorization & Hirano et al.~\cite{hirano2023Adversarial} & Zhu et al.~\cite{zhu2025Securing}; Banerjee et al.~\cite{banerjee2024Fine}; Konstantinidis et al.~\cite{konstantinidis2024Finllama} \\
R2I & Reg-to-Integrity & Faysal et al.~\cite{faysal2026Agentic} & Shukanayev~\cite{shukanayev2025Who}; Hornuf et al.~\cite{hornuf2025Making}; Bain and Subirana~\cite{bain2003Legalising} \\
\bottomrule
\end{tabular}
\end{table*}

\noindent Vectors with fewer released direct-support papers (R2I, M2A, C2E) are more speculative; we mark these explicitly throughout \S\ref{sec:comparison} and flag them as priorities for future empirical work.

\subsection{Cross-Layer Vector Derivation}

The 12 cross-layer attack vectors in \S\ref{sec:comparison} were derived through a structured three-step process. \textbf{Step 1 (Layer identification)}: we enumerated the main layers of autonomous financial agent systems, including reasoning, tools, custody, inter-agent protocols, settlement, oracles, identity, and compliance. \textbf{Step 2 (Pairwise analysis)}: for each ordered layer pair $(L_i, L_j)$ we asked whether compromise at $L_i$ could induce harm at $L_j$ while bypassing $L_j$'s own defenses. \textbf{Step 3 (Consolidation)}: attack paths with direct evidence were retained, while recurring but previously unnamed paths were generalized into cross-layer vectors.

Historically, this derivation also benefits from older agent-commerce literature that predates LLMs but already exposed the relevant trust, delegation, and protocol-design problems. Early platform-mediated agents and web-commerce systems established the centralized trust model~\cite{nwana1998Agent, moukas2000Agent}; later multi-agent finance work explored negotiation, contract-net coordination, and reputation attribution in ways that remain directly relevant to evaluator-mediated and protocol-mediated commerce today~\cite{sierra2001Agent, gateau2004Contract, wang2008Attributions}. Programmatic-agent research also made the trade-off between predictability and flexibility explicit well before modern LLM agents~\cite{lee2002Intelligent, hua2001Agents, ma1999Agents, redmond2002The, murdock2002The}.

\subsection{Analysis Dimensions}

Our review uses two complementary coding layers. First, each retained work is assigned to a primary topical bucket, with an optional secondary bucket when a work substantively bridges communities: \textbf{C0} = background, legal, benchmark, and framing sources that inform the review but do not fit cleanly into one of the five technical communities; \textbf{C1} = LLM security and alignment; \textbf{C2} = autonomous agent architectures; \textbf{C3} = blockchain and DeFi security; \textbf{C4} = financial technology and algorithmic trading; and \textbf{C5} = multi-agent systems and mechanism design. Second, the SoK synthesis itself is organized around the paper's five security dimensions: \textbf{D1} = agent integrity, \textbf{D2} = transaction authorization, \textbf{D3} = inter-agent trust, \textbf{D4} = market manipulation, and \textbf{D5} = regulatory compliance. Each retained work receives one primary synthesis-dimension assignment and, when the work materially spans multiple parts of the framework, additional dimension flags.

Our final public corpus spans 1994--2026. We use topical buckets (\textbf{C0}--\textbf{C5}) to track which research communities each source comes from, and synthesis dimensions (\textbf{D1}--\textbf{D5}) to organize how each source contributes to the security framework.

\subsection{Positioning Against Related LLM-Agent Security Surveys}
\label{sec:positioning}

Recent LLM-agent surveys cover prompt injection, tool misuse, multi-agent trust, and runtime control in domain-general settings~\cite{zhan2025UnifiedThreat, he2025TrustAgent, multiagent2025Security}. Adjacent survey and systems work spans commerce-oriented agentic AI adoption~\cite{balaskas2026Delegation, brohi2025Landscape}, zero-trust and cross-domain agent security~\cite{zerotrust2025MultiLLM, crossdomain2025MultiAgent}, communication- and protocol-layer defenses for agent networks~\cite{agentcommsec2025Communication, agentprotocolsec2025ACP}, IAM or trust-fabric style authorization layers~\cite{trustfabric2025Formal, saga2025Tokenized}, and architectural views of an on-chain agent economy~\cite{agenteconomy2026Architecture}. Our scope is narrower but operationally deeper: we focus on financial irreversibility, custody, settlement integrity, evaluator-mediated commerce, market manipulation, and compliance exposure, and we connect these threats directly to deployed or emerging agent-payment protocols such as ERC-8004, AP2, OKX APP, x402, MPP, and ACP.

This finance-specific framing also changes how otherwise generic controls are interpreted. Safety benchmarks such as RiskyBench and quit-style loss-limiting behavior map naturally to authorization and circuit-breaker questions in financial agents~\cite{riskybench2026Bench, quit2025Quitting}. Deterministic pre-action policy engines such as ACP-Control and OAP can gate tool use before execution~\cite{acpcontrol2026Admission, oap2026Auth}, while inter-agent trust taxonomies and wallet- and credential-security analyses clarify which assumptions belong to identity, stake, proof, or custody rather than to the LLM alone~\cite{interagenttrust2025Taxonomy, walletsok2023Erinle}.

\section{Systematization Framework}
\label{sec:framework}

\begin{figure}[t]
\centering
\includegraphics[width=0.9\linewidth]{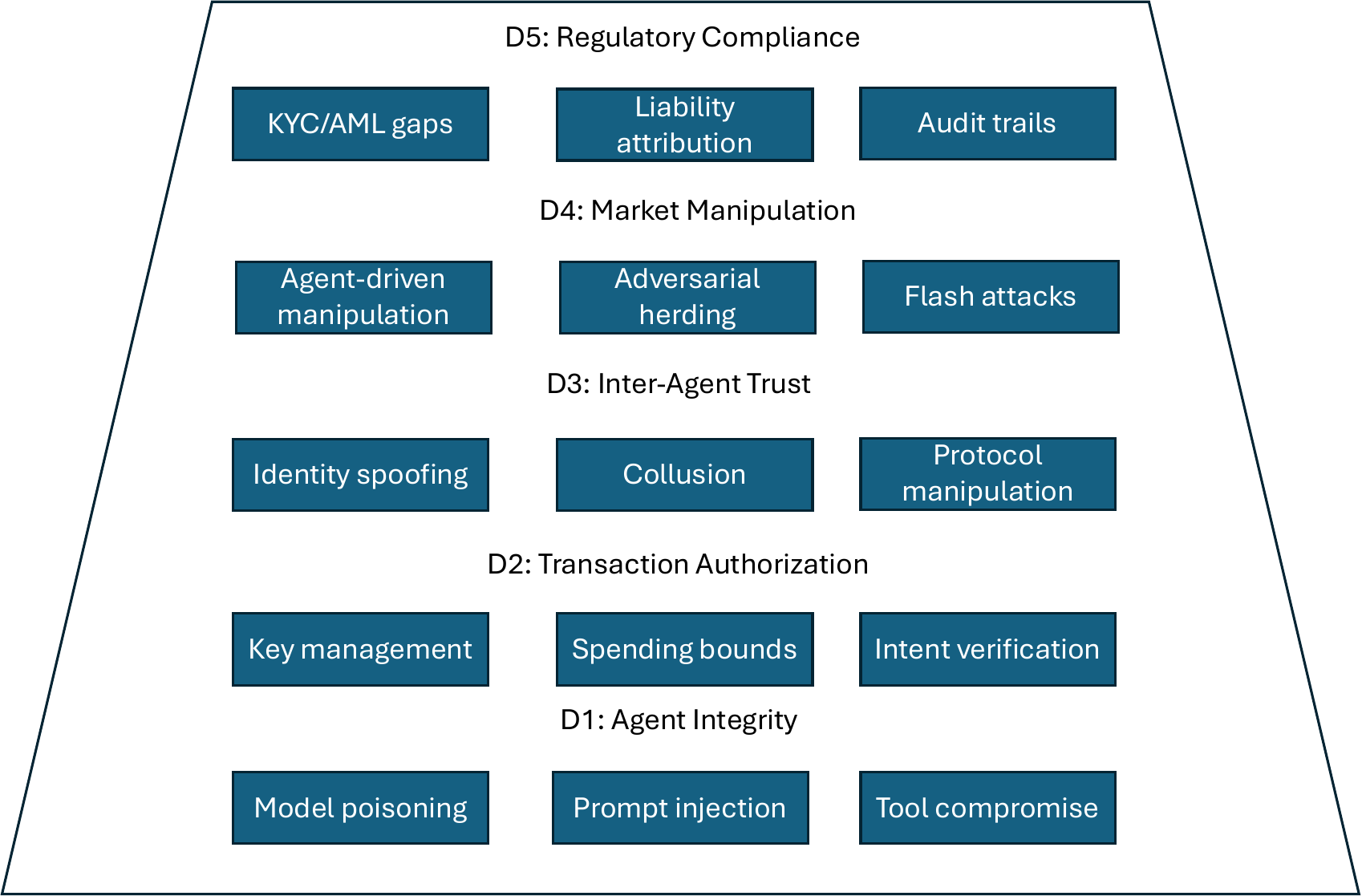}
\caption{Five-dimensional threat taxonomy for autonomous financial agents.}
\label{fig:taxonomy}
\end{figure}

We propose a five-dimensional threat taxonomy for autonomous financial agents, illustrated in Figure~\ref{fig:taxonomy}. Each dimension captures a distinct category of security concern that arises from the intersection of LLM-based autonomy and financial operations.

\subsection{Dimension 1: Agent Integrity}

Agent integrity concerns whether the agent's decision-making process has been compromised, causing it to deviate from its intended financial objectives. This dimension is unique to LLM-based agents because the natural language interface that enables their flexibility also creates attack surfaces that do not exist in traditional automated systems.

\subsubsection{Prompt Injection Attacks}
Prompt injection remains the most direct threat to agent integrity~\cite{greshake2023Indirect, liu2024Automatic, preprint2026Security}. Recent work shows that these attacks can be optimized automatically at small data budgets and can still be difficult to reconstruct cleanly after the fact when investigating compromised agent workflows~\cite{liu2024Automatic, chernyshev2024Forensic}. In financial contexts, prompt injection takes on heightened severity because successful attacks can trigger irreversible financial transactions. We identify three categories of prompt injection specific to financial agents:

\emph{Direct injection via data feeds.} Financial agents consume market data, news feeds, and social media signals as inputs to their decision-making. An adversary can embed malicious instructions in these data sources. For example, a crafted news article or social media post containing hidden prompt injection text could instruct an agent to execute a specific trade~\cite{greshake2023Indirect}. Unlike generic prompt injection, financial injection can be \emph{economically motivated} because the attacker profits from the manipulated trade.

\emph{Injection through smart contract metadata.} On-chain data, including token names, contract descriptions, and transaction memos, can carry prompt injection payloads. When an agent reads on-chain data to inform its decisions, these payloads can alter its behavior~\cite{acharya2025Secure}. The Ethereum naming system (ENS) and token metadata fields are particularly vulnerable vectors.

\emph{Injection via inter-agent communication.} In multi-agent settings, one agent's output becomes another agent's input. A compromised or adversarial agent can embed prompt injection attacks in its negotiation messages, trade proposals, or status updates~\cite{allouah2025What, zhang2025Guruagents}. This creates the possibility of \emph{cascading compromises} where a single compromised agent subverts an entire network of agents.

\subsubsection{Model Poisoning and Backdoors}
Financial agents that undergo fine-tuning on financial data~\cite{zhu2025Securing, banerjee2024Fine, konstantinidis2024Finllama} are vulnerable to data poisoning attacks. An adversary who can influence the training data can embed backdoor triggers that cause specific financial behaviors (e.g., always buying a particular token when a certain phrase appears in market data). The challenge is amplified by the opacity of LLM decision-making: a poisoned agent may perform normally on most inputs while executing adversarial trades on trigger inputs~\cite{hirano2023Adversarial}.

\subsubsection{Tool and Plugin Compromise}
Autonomous financial agents rely on external tools accessed through protocols like MCP~\cite{research2024Model, security2025Model} for market data, trade execution, and portfolio analysis. A compromised tool can feed the agent false information (e.g., incorrect prices) or execute transactions that differ from what the agent requested. The security of the agent is therefore bounded by the security of its weakest tool integration~\cite{security2025Model, acharya2025Secure}. Recent defense proposals therefore attempt to attribute tool invocations back to the causally responsible prompt context so that suspicious calls can be blocked before execution~\cite{attriguard2026Causal}.

\subsubsection{Memory Injection and Persistent Store Poisoning}
Long-running financial agents maintain persistent memory stores, such as vector databases, key-value caches, or episodic memory logs, that inform future reasoning. \emph{Memory injection} attacks corrupt this persistent state to influence the agent's future decisions without requiring continuous prompt injection~\cite{zhan2025UnifiedThreat, memoryinject2025Web3, memorygraft2025Persistent, memrobtrust2025Robustness}. An adversary who can write to a memory store (e.g., by submitting a malicious transaction whose metadata is indexed, or by injecting content that the agent retrieves via RAG) can implant false ``learned experiences'' that bias the agent toward adversary-preferred trades in future sessions. Unlike prompt injection, which acts in the current context window, memory injection persists across sessions and may be difficult to detect because the agent's behavior changes gradually.

We extend the D1 taxonomy to include three memory-specific attack sub-types: (1) \emph{Write-path poisoning}, in which an adversary injects adversarial content through any channel that the agent records to memory (on-chain data ingestion, inter-agent messages, tool outputs); mitigation requires write-access controls that restrict which data sources can populate the agent's long-term memory, ideally enforced by content provenance tags analogous to inter-agent origin tagging (see \S\ref{sec:comparison}); (2) \emph{Retrieval manipulation}, in which adversarially crafted embeddings can cause malicious memories to be preferentially retrieved for target query contexts if the agent uses semantic search (e.g., a vector database) for memory retrieval; mitigation requires adversarial-robustness testing of the embedding model and retrieval pipeline; (3) \emph{Memory staleness exploitation}, in which out-of-date memories can cause agents to act on superseded policy parameters or market conditions; mitigation requires TTL-bounded memory entries with mandatory refresh for high-value decisions. Defense: cryptographic provenance tags on all memory writes recording the source, timestamp, and a hash of the originating context; memory verifiers that reject writes from untrusted sources; periodic memory audits comparing stored experiences against independently verifiable on-chain records. Recent multi-agent designs also explore privacy-preserving shared-memory layers with explicit trust weighting as a mitigation direction, rather than treating memory as an opaque global scratchpad~\cite{superlocalmem2026SuperLocal}.

\subsection{Dimension 2: Transaction Authorization}

Transaction authorization concerns whether the agent's financial actions are properly bounded and verifiable. This dimension addresses the fundamental question: \emph{how do we ensure that an autonomous agent only executes transactions it is authorized to perform?}

\subsubsection{Credential and Key Management for Autonomous Agents}
When an agent holds digital-wallet credentials, payment account secrets, or cryptographic private keys, as in OpenClaw's architecture~\cite{steinberger2025Openclaw}, credential security becomes critical. Traditional account and key management assumes a human user who can recognize and resist social engineering attacks. An LLM agent, by contrast, can be prompt-injected into revealing or misusing credentials~\cite{acharya2025Secure}. Multi-signature schemes, scoped API tokens, and hardware security modules (HSMs) can limit exposure, but they introduce latency that may be incompatible with high-frequency financial operations.

\subsubsection{Spending Policies and Bounds}
ERC-8004~\cite{goenka2026Tesspay} introduces on-chain spending limits and allowance mechanisms for agent-controlled wallets. However, the granularity of these policies is a design challenge. Overly permissive policies enable unauthorized large transactions; overly restrictive policies impede legitimate agent operations. Prior work on agent authorization in enterprise systems~\cite{papazoglou2001Agent, kerr2001Ensuring} provides relevant frameworks, but these must be adapted for the adversarial environment of public blockchains. Foundational capability-oriented delegation mechanisms, from macaroons and object-capability designs to newer semantic task-to-scope matching and workflow-scoped agent credentials, point to the same design rule: agent permissions should be attenuated, contextual, and bound to a narrow task rather than to a standing broad wallet or account authority~\cite{birgisson2014Macaroons, miller2006Robust, helou2025DelegatedAuthorization, priorajwt2025Workflow}.

\subsubsection{Intent Verification}
A key challenge in agent-mediated finance is verifying that the agent's \emph{intent}, as formed by LLM reasoning, matches the \emph{action}, as encoded in a payment instruction or blockchain transaction~\cite{acharya2025Secure}. An agent may reason correctly about a trade but produce an incorrect transaction due to encoding errors, unit conversion mistakes, or manipulation of intermediate representations. AP2's payment intent mechanism~\cite{liu2026Agenticpay} partially addresses this by separating intent declaration from execution, enabling pre-execution verification. MPP adds a complementary transport-layer safeguard: servers can bind a payment challenge to a specific request body using content digests, reducing the risk that a valid credential is replayed against a modified API call~\cite{mpp2026Protocol}.

\subsection{Dimension 3: Inter-Agent Trust}

As the agentic economy grows, agents increasingly interact with other agents rather than with humans~\cite{birch2025Agentic, zhang2026Agentic, virtuals2026ACP}. This introduces trust challenges that have no direct parallel in human-operated systems.

\subsubsection{Agent Identity and Authentication}
In a multi-agent marketplace, how does one agent verify the identity and trustworthiness of another? Current systems often rely on cryptographic signatures or account-bound credentials, especially wallet addresses in on-chain settings, but these do not convey information about the agent's capabilities, authorization level, or operating policies~\cite{acharya2025Secure, shukanayev2025Who}. An adversarial agent can create multiple identities (Sybil attack) to manipulate reputation systems or conduct wash trading~\cite{ping2005Deceit}.

\subsubsection{Negotiation Integrity}
Agent-to-agent negotiation, which is a core function in agentic commerce~\cite{sandholm1999Automated, sandholm2000Agents, chung2000NegotiationModel, de2001Bilateral}, is vulnerable to manipulation when one party is an LLM. Traditional negotiation protocols assume rational, self-interested actors; LLM agents may be susceptible to persuasion, anchoring effects, or adversarial framing that exploits their language understanding~\cite{allouah2025What, kapoor2025Build}. An adversary can craft negotiation messages that exploit LLM biases to obtain unfavorable terms for the victim agent.

\subsubsection{Collusion and Coordination Attacks}
Multiple adversarial agents can coordinate to manipulate market prices, conduct pump-and-dump schemes, or corner markets~\cite{wang2025Modeling, moreno2025Predicting}. The ability of LLM agents to communicate in natural language makes collusion harder to detect than in traditional algorithmic settings, where coordination requires explicit protocol-level mechanisms. Multi-agent systems research has studied coalition formation~\cite{sandholm2000Agents, solodukha2010Multi} but not in the context of LLM-driven strategic behavior.

\subsubsection{Cascading Compromise via Prompt Infection}
A distinctive threat in LLM-based multi-agent networks is \emph{self-replicating prompt injection}~\cite{lee2024PromptInfection}, where a single compromised agent embeds adversarial instructions in its outgoing messages that cause recipient agents to replicate the injection onward. Prompt Infection~\cite{lee2024PromptInfection} demonstrates experimentally that such infections can propagate through multi-agent networks from a single entry point, analogous to biological contagion. In a financial agent network, this mechanism could disseminate malicious trading instructions across the network before any single agent's safety checks trigger, causing coordinated unauthorized transactions at scale. We therefore recommend incorporating inter-agent message origin tagging and stricter sanitization of agent-sourced content as baseline security requirements for multi-agent financial deployments.

\subsection{Dimension 4: Market Manipulation}

Autonomous agents introduce novel market manipulation risks that extend beyond traditional concerns about algorithmic manipulation.

\subsubsection{Agent-Driven Market Manipulation}
An adversary who controls one or more autonomous agents can use them to conduct market manipulation at machine speed~\cite{xiao2025TradingWith, guo2025Findebate, byrd2025PumpDump}. Unlike human manipulators, agent-based schemes can operate continuously, adapt to market responses in real time, and coordinate across multiple markets simultaneously. Recent evidence also suggests that agentic trading systems can drift into manipulation-like behavior even when profit maximization, rather than explicit market abuse, is the nominal objective~\cite{byrd2025PumpDump}. The combination of LLM reasoning (for strategy adaptation) and automated execution (for speed) creates a more potent manipulation capability than either alone.

\subsubsection{Adversarial Herding}
Because many LLM agents are built on similar foundation models and trained on overlapping data, they may exhibit correlated behavior, a form of ``herding'' that can amplify market movements~\cite{nabar2023Conservative}. An adversary who understands these correlations can craft market signals (e.g., fake news, manipulated sentiment indicators) designed to trigger correlated responses across multiple agents, causing flash crashes or artificial price spikes~\cite{hirano2023Adversarial, kim2023Gans}.

A critical property of adversarial herding is that per-agent authorization controls cannot prevent it: each agent's individual action may be within its authorized scope, yet the aggregate effect is harmful. Mitigations must therefore operate at the market and regulatory levels, including portfolio-level circuit breakers, model diversity mandates for large agent fleets, and market-wide circuit breakers at the exchange or protocol level~\cite{nabar2023Conservative}.

\subsubsection{Sandwich Attacks on Agent Transactions}
In DeFi settings, autonomous agents that broadcast pending transactions are vulnerable to sandwich attacks, where an adversary front-runs the agent's transaction to manipulate the price and then back-runs to profit from the price impact~\cite{assis2024Analysis}. While sandwich attacks exist in traditional DeFi, autonomous agents are particularly vulnerable because they may lack the real-time monitoring and evasive capabilities of specialized MEV (Maximal Extractable Value) bots. Established DeFi-native mitigations that autonomous agents can adopt include: \emph{private order flow} via MEV-protected RPC endpoints (e.g., Flashbots Protect, MEV Blocker) that route transactions directly to block builders without public mempool exposure; \emph{orderflow auctions} (OFAs) that allow agents to auction exclusive transaction rights to searchers who return MEV rebates rather than extracting them adversarially; and \emph{slippage-bounded transactions} that set tight maximum acceptable price impact, causing the transaction to revert if a sandwich attack inflates cost beyond the bound. For autonomous agents, the key implementation challenge is incorporating these defenses into the agent's transaction submission pipeline without introducing latency that degrades strategy performance. We provide concrete trade-off guidance: \emph{private RPC endpoints} (Flashbots Protect, MEV Blocker) add zero submission latency versus public RPC but introduce a routing delay of 50--200\,ms to the next block inclusion due to builder propagation; for strategies where execution timing within a block is non-critical (e.g., DCA, rebalancing), this is acceptable. \emph{OFAs} add a 300--800\,ms auction window before the transaction is committed to a builder, making them unsuitable for latency-sensitive arbitrage but appropriate for large, price-impact-sensitive trades where MEV rebates offset the latency cost. \emph{Slippage bounds} add zero latency but require careful calibration: too tight a bound causes excessive transaction reverts in volatile markets; Careful calibration of slippage bounds is required to balance execution success rate against adversarial profitability.

\subsection{Dimension 5: Regulatory Compliance}

The deployment of fully autonomous financial agents raises profound regulatory questions that have direct security implications.

\subsubsection{KYC/AML for Non-Human Actors}
Current Know Your Customer (KYC) and Anti-Money Laundering (AML) frameworks assume human account holders~\cite{hornuf2025Making, shukanayev2025Who}. When an autonomous agent controls a digital wallet or payment account and transacts independently, who is the ``customer''? The agent's deployer? The model provider? The framework developer? This ambiguity can be exploited by adversaries to launder money through chains of agent-mediated transactions that obscure the ultimate beneficial owner~\cite{faysal2026Agentic, hacini2025Llm}.

In the United States, FinCEN's Bank Secrecy Act (BSA) regulations require money services businesses (MSBs) to file Currency Transaction Reports (CTRs) for cash transactions exceeding \$10,000 and Suspicious Activity Reports (SARs) for suspicious transactions of \$5,000 or more~\cite{fincen2019AML}. Autonomous agents conducting financial transactions at scale could constitute unregistered MSBs, and the N2C attack vector (\S\ref{sec:comparison}) specifically exploits BSA structuring prohibitions. In the European Union, the Markets in Crypto-Assets (MiCA) Regulation (Regulation 2023/1114) establishes licensing requirements for crypto-asset service providers that would apply to platforms hosting autonomous agent crypto-wallets~\cite{mica2023Regulation}. The FATF's 2021 and 2023 guidance on Virtual Assets and Virtual Asset Service Providers (VASPs) explicitly addresses algorithmic entities and requires member states to extend VASP AML obligations to entities that facilitate VA transfers on behalf of customers~\cite{fatf2023VASP}, a definition that arguably encompasses autonomous agent infrastructure operators. Mapping these specific obligations to our taxonomy: D5 (regulatory compliance) intersects with D2 (transaction authorization) through SAR filing requirements that would mandate real-time detection of structuring patterns (N2C attacks) and with D3 (inter-agent trust) through VASP travel rule requirements for transmitting beneficiary information in agent-to-agent payments.

\subsubsection{VASP Classification and Principal Liability}
A fundamental compliance question is whether an autonomous agent itself constitutes a Virtual Asset Service Provider (VASP) or whether its deployer bears VASP obligations. FATF Guidance (2023)~\cite{fatf2023VASP} defines a VASP as an entity that \emph{conducts} VA transfers as a business on behalf of others. Under this definition, the \emph{infrastructure operator}, not the individual agent instance, is the liable VASP, bearing KYC/AML obligations for all agents on their platform. Practically, this implies a three-tier principal model: (a)~the \emph{deployer} must complete KYC at agent deployment time, binding their legal identity to the agent's DID; (b)~the \emph{infrastructure operator} maintains the AML monitoring stack (N2C detectors, CTR/SAR pipelines); (c)~the \emph{agent instance} is an automated execution process, not itself a regulated entity. This model avoids regulatory uncertainty: no jurisdiction currently licenses AI agents as financial institutions.

\subsubsection{Liability Attribution}
When an autonomous agent causes financial harm through a security breach, market manipulation, or erroneous trade, determining liability is challenging~\cite{stocker2025PrincipalAgent, bain2003Legalising}. The multi-layered architecture of agent systems (model provider, framework developer, tool provider, deployer) creates a diffusion of responsibility that adversaries can exploit. Without clear liability frameworks, there is insufficient incentive for any single party to invest in comprehensive security~\cite{shukanayev2025Who, hornuf2025Making}.

\subsubsection{Audit Trail Requirements}
Financial regulations typically require detailed audit trails of all transactions~\cite{dagostino2024Capturing}. For autonomous agents, this requires logging not just the transactions themselves but the LLM reasoning that led to them. Current LLM architectures make this challenging: the mapping from input context to output action is opaque, and agents may process thousands of data points to arrive at a single trading decision. The x402 protocol~\cite{goenka2026Tesspay} provides some audit trail capabilities through its payment metadata, and MPP extends this with explicit challenges, credentials, receipts, and optional request-body digests~\cite{mpp2026Protocol}; however, these protocol traces are still insufficient for full regulatory compliance.

\subsubsection{Obligation-to-Control Mapping}
Financial-agent compliance translates abstract obligations into engineering controls. In practice, AML/CFT and market-abuse requirements imply beneficial-owner binding, cumulative exposure monitoring, Travel Rule style provenance exchange for agent-to-agent transfers, and tamper-evident audit logs~\cite{hornuf2025Making, faysal2026Agentic, dagostino2024Capturing}. This makes D5 inseparable from the rest of the framework: authorization limits help block structuring, identity and settlement metadata support provenance exchange, and integrity-protected logs are needed when agent decisions are later audited.

\subsection{Incident Lessons}

Representative incidents and constructed scenarios converge on three lessons. First, attacks are usually cross-dimensional: token-metadata injection links D1 to D2, compromised tools link reasoning to market harm, and evaluator manipulation links D3 to settlement. Second, platform-level failures can be systemic, as illustrated by the Virtuals launch vulnerability, where a single infrastructure flaw could have affected every agent using the same launch path~\cite{shittu2025VirtualsProtocolBug}. Third, the most damaging campaigns are often cumulative rather than spectacular: subtle oracle drift, repeated negotiation exploitation, or under-threshold structuring can remain locally plausible while producing significant portfolio or compliance harm over time. Lifecycle-oriented analysis of OpenClaw reinforces this point by showing how initialization, memory, inference, and execution stages create compounding attack opportunities rather than isolated bugs~\cite{openclaw2026Lifecycle}. These observations motivate the layered controls summarized in \S\ref{sec:trends}.

\section{Comparative Analysis}
\label{sec:comparison}

In this section, we compare existing approaches to securing autonomous financial agents across the dimensions of our taxonomy, analyze their trade-offs, and identify cross-layer attack vectors.

\subsection{Defense Approaches by Dimension}

\subsubsection{Agent Integrity Defenses}

Table~\ref{tab:integrity_defenses} summarizes the principal defense categories for agent integrity.

\begin{table}[t]
\centering
\footnotesize
\caption{Comparison of agent integrity defense approaches.}
\label{tab:integrity_defenses}
{\renewcommand{\arraystretch}{1.06}
\begin{tabularx}{\columnwidth}{L{1.8cm}Y L{1.8cm}C{1.15cm}}
\toprule
\textbf{Approach} & \textbf{Mechanism} & \textbf{Coverage} & \textbf{Overhead} \\
\midrule
Input sanitization & Filter malicious prompts from data feeds before model ingestion & Direct injection & \levelLow \\
Instruction hierarchy & Privilege separation between system and user/data prompts & Direct \& indirect injection & \levelMed \\
Output validation & Verify proposed actions against policy before execution & All integrity threats & \levelHigh \\
Redundant reasoning & Cross-check decisions with multiple independent LLM instances & Model poisoning; subtle injection & \levelVHigh \\
Formal verification & Prove bounded safety properties of the agent pipeline & Broad but partial & Infeasible \\
\bottomrule
\end{tabularx}}
\end{table}

\emph{Input sanitization}~\cite{greshake2023Indirect} is the most straightforward defense, filtering potentially malicious content from data feeds before they reach the agent's context. However, in financial applications, aggressive filtering risks removing legitimate market signals. A news headline about a ``crash'' might be filtered as a potential injection vector when it is in fact critical market information.

\emph{Instruction hierarchy} approaches~\cite{greshake2023Indirect, allouah2025What} establish privilege levels where system-level instructions (e.g., ``never transfer more than 1 ETH per transaction'') cannot be overridden by data-level content. While effective against many injection attacks, these approaches face challenges when agents must reason about user-provided financial objectives that necessarily interact with system constraints.

\emph{Output validation}~\cite{acharya2025Secure} interposes a verification layer between the agent's reasoning and its actions, checking proposed transactions against policy constraints before execution. This is the most robust single defense but introduces latency that can be costly in fast-moving financial markets. The validation layer itself must be secured against bypass~\cite{security2025Model}.

\emph{Redundant reasoning}~\cite{yu2024Fincon, zhang2025Guruagents} uses multiple independent LLM instances to cross-check financial decisions, similar to multi-factor authentication but applied to AI reasoning. While effective at catching individual model failures, this approach multiplies computational costs and still fails if all instances share the same vulnerability (e.g., a common training data bias).

\emph{Runtime verification and capability bounding.} Financial agents benefit from an independent control layer between reasoning and tool/action execution. ZTRV-style checks validate that each action remains bound to the current workflow context before execution~\cite{ztrv2026Runtime}, while Agent-Sentry constrains action sequences using execution provenance and capability graphs~\cite{agentsentry2026Runtime}. These mechanisms complement output validation and payment authorization: the runtime verifier can reject replayed or context-drifted actions before they reach custody, while the custody layer still enforces transaction scope. Independent reproduction in financial agentic settings (with irreversible on-chain transactions and financial-specific attack scenarios) remains an open experimental challenge that we identify in \S\ref{sec:future}. These controls complement the Layer 1 (prompt hardening) and Layer 2 (reasoning verification) proposals in our defense architecture (\S\ref{sec:trends}).

\emph{Tool selection integrity.} The tool-\emph{selection} stage is an independent attack surface: compromised registries, misleading tool descriptions, or forged provenance metadata can redirect an agent toward malicious tools before any invocation occurs. This motivates pre-invocation verification of tool provenance and description integrity, not only post-invocation output validation.

\subsubsection{Transaction Authorization Defenses}

The design space for transaction authorization defenses spans a spectrum from fully on-chain enforcement to fully off-chain policy engines.

\emph{Smart contract guardrails.} ERC-8004~\cite{goenka2026Tesspay} enables on-chain enforcement of spending limits, per-transaction caps, and time-locked operations. These guardrails are tamper-resistant (enforced by consensus) but inflexible because modifying policies requires on-chain transactions with associated gas costs and latency. Recent work has explored programmable spending policies that combine on-chain enforcement with off-chain configuration~\cite{goenka2026Tesspay}.

\emph{Multi-signature and threshold schemes.} Requiring multiple signatures for high-value transactions provides strong authorization guarantees~\cite{acharya2025Secure}. In multi-agent settings, this can be implemented as requiring agreement among multiple independent agents before executing a trade. However, this approach assumes that the multiple signers are truly independent; if they share the same LLM backbone, a universal attack might compromise all of them simultaneously.

\emph{Intent-action verification.} AP2's payment intent mechanism~\cite{liu2026Agenticpay} enables pre-execution verification by separating the declaration of intent from its execution. A verifier can confirm that the intended action matches the proposed transaction before it is submitted to the blockchain. This approach is particularly valuable for complex transactions involving multiple steps or cross-chain operations. MPP provides a related transport-layer defense through challenge--credential--receipt verification and digest-bound requests, allowing servers to ensure that the paid request is the same request that is ultimately executed~\cite{mpp2026Protocol}.

\subsubsection{Inter-Agent Trust Defenses}

\emph{Decentralized identity (DID).} Agent identity can be anchored in decentralized identity systems that provide verifiable credentials about an agent's capabilities, authorization level, and operating history~\cite{acharya2025Secure}. However, DID systems currently lack standardized credential types for autonomous agents, and the process of issuing credentials for non-human entities raises unresolved governance questions.

\emph{Reputation systems.} On-chain reputation systems track agents' transaction histories and compute trust scores~\cite{komiak2004Understanding, wang2007Recommendation}. These systems face the cold-start problem (new agents have no reputation) and are vulnerable to reputation farming and wash trading by adversarial agents~\cite{ping2005Deceit}.

\emph{Escrow and atomic settlement.} Payment protocols like AP2 support escrow mechanisms where funds are locked in a smart contract until both parties confirm transaction completion~\cite{liu2026Agenticpay}. Virtuals Protocol's ACP extends this with an evaluator agent model, where a third-party agent assesses deliverable quality before releasing escrow funds~\cite{virtuals2026ACP}. While this reduces the need for mutual trust between transacting agents, it introduces a new trust assumption on the evaluator because a compromised evaluator can systematically approve fraudulent deliverables or reject legitimate ones, enabling the C2E attack vector described in \S\ref{sec:comparison}. ERC-8183 formalizes this pattern with on-chain state machines and extensible hooks~\cite{virtuals2026ERC8183}.

The security of evaluator agents in ACP/ERC-8183 warrants explicit threat modeling. We identify four concrete evaluator attack scenarios: (1) \emph{direct bribery}, where a provider agent compensates an evaluator out-of-band to approve a fraudulent deliverable; (2) \emph{Sybil evaluator clusters}, where an attacker deploys many evaluator identities to influence the evaluator selection pool; (3) \emph{evaluator--provider collusion}, where the same controlling party operates both roles and systematically manipulates escrow outcomes; and (4) \emph{adversarial evaluator substitution}, where an attacker front-runs evaluator-assignment transactions on-chain to insert a malicious evaluator. Mitigations include: bonded evaluators with slashing for misconduct; Byzantine fault-tolerant committee sizing with VRF-based selection to prevent front-running; TEE-backed cryptographic independence attestation; and on-chain statistical anomaly monitoring to flag evaluators with abnormal approval patterns~\cite{virtuals2026ERC8183}. Atomic settlement ensures that multi-step transactions either complete entirely or revert entirely, preventing partial execution attacks.

\subsection{Cross-Layer Attack Vectors}

A critical finding of our analysis is that the most dangerous attacks on autonomous financial agents exploit \emph{cross-layer interactions}, where a vulnerability at one layer triggers a cascading failure at another. We identify and characterize all 12 cross-layer attack vectors below; Table~\ref{tab:crosslayer} provides a concise overview with adversary preconditions and layer paths.

Three distinctions matter most operationally. \emph{T2R vs.\ T2T}: T2R corrupts reasoning through false data, while T2T corrupts execution after correct reasoning; the former is mitigated by provenance checks and cross-validation, the latter by end-to-end intent binding. \emph{T2R vs.\ O2P}: T2R is usually acute and transaction-local, whereas O2P is chronic and cumulative, requiring longitudinal monitoring rather than single-trade anomaly detection. \emph{P2T vs.\ P2K}: P2T induces a new unauthorized action; P2K coerces signing itself and therefore requires a hard separation between cognition and custody.

Across the 12 vectors, the most immediate deployment risks are P2T, T2R, T2T, and S2I because they convert public inputs, tools, or dependencies into directly authorized financial actions~\cite{greshake2023Indirect, security2025Model}. C2E, O2P, and N2C are slower-burn but often harder to detect because harm accumulates over time. R2I and M2A remain more speculative in the current corpus and should be treated as early-warning categories rather than equally grounded threats.

Recent protocol and deployment analyses sharpen these distinctions. MCP-specific studies point to capability-attestation gaps, unsafe trust propagation, and over-privileged tool wrappers as concrete precursors to T2R-style failure~\cite{mcpanalysis2026MCP, agentaudit2026Audit}. Supply-chain exploitation work likewise shows that poisoned dependencies and prompt templates can bypass otherwise sound reasoning-layer defenses, which is why S2I belongs in the top deployment tier rather than being treated as a generic software risk~\cite{ruan2025PracticalExploit}.

\begin{table}[t]
\centering
\footnotesize
\caption{Summary of 12 cross-layer attack vectors for autonomous financial agents.}
\label{tab:crosslayer}
{\renewcommand{\arraystretch}{1.05}
\begin{tabularx}{\columnwidth}{L{0.7cm}L{1.55cm}L{1.7cm}Y}
\toprule
\textbf{ID} & \textbf{Name} & \textbf{Layer Path} & \textbf{Core Mechanism} \\
\midrule
\texttt{P2T} & Prompt-to-Transaction & LLM $\to$ Blockchain & Injected prompt triggers signed tx \\
\texttt{T2R} & Tool-to-Reasoning & Tool $\to$ Reasoning & False data poisons decision \\
\texttt{A2M} & Agent-to-Market & Inter-agent $\to$ Market & LLM bias exploited in negotiation \\
\texttt{R2I} & Reg-to-Integrity & Compliance $\to$ Market & Regulatory gap enables laundering \\
\texttt{T2T} & Tool-to-Transaction & Tool $\to$ Blockchain & Tool modifies tx params post-reasoning \\
\texttt{P2K} & Prompt-to-Key & LLM $\to$ Custody & Injection bypasses key custody boundary \\
\texttt{M2A} & Model-to-Authorization & Model $\to$ Authorization & Backdoor defeats spending policy check \\
\texttt{C2E} & Collusion-to-Escrow & Multi-agent $\to$ Settlement & Colluding evaluators drain escrow \\
\texttt{O2P} & Oracle-to-Position & Oracle $\to$ Portfolio & Cumulative drift via subtle feed manipulation \\
\texttt{I2M} & Identity-to-Market & Reputation $\to$ Market & Sybil trust enables coordinated manipulation \\
\texttt{N2C} & Neg-to-Compliance & Protocol $\to$ Compliance & Structuring payments evades AML threshold \\
\texttt{S2I} & Supply-to-Integrity & Supply chain $\to$ All & Backdoored plugin silently alters transactions \\
\bottomrule
\end{tabularx}}
\end{table}

\subsection{Comparative Assessment of Protocols and Interfaces}

Table~\ref{tab:protocols} compares representative protocols and execution interfaces used by autonomous financial agents.

\begin{table*}[t]
\centering
\small
\caption{Security property comparison of representative agent-commerce protocols and execution interfaces. `N/A' denotes a dimension outside the artifact's intended scope.}
\label{tab:protocols}
\resizebox{\textwidth}{!}{%
\begin{tabular}{lccccc}
\toprule
\textbf{Protocol / Interface} & \textbf{Agent Integrity} & \textbf{Transaction Authorization} & \textbf{Inter-Agent Trust} & \textbf{Market Manipulation} & \textbf{Regulatory Compliance} \\
\midrule
Virtuals/ACP~\cite{virtuals2026ACP} & GAME framework & Escrow settlement & Evaluator agents & Token caps & On-chain logs \\
ERC-8183~\cite{virtuals2026ERC8183} & N/A (standard) & Job state machine & 3-role trust model & Hook-based & Reputation \\
ERC-8004~\cite{goenka2026Tesspay} & N/A (protocol level) & On-chain guardrails & Token-based ID & Spending caps & Audit logs \\
AP2~\cite{liu2026Agenticpay} & N/A (protocol level) & Intent verification & Payment attestation & N/A & Payment logs \\
OKX APP~\cite{okx2026AgentPaymentsProtocol, okx2026AgentPaymentsProtocolDocs} & N/A (protocol level) & Challenge/credential auth & A2A escrow/dispute & N/A & Status queries \\
x402~\cite{goenka2026Tesspay} & N/A (protocol level) & Per-request auth & HTTP-level auth & Rate limiting & Request logs \\
MPP~\cite{tempo2026MachinePayments, mpp2026Protocol} & N/A (protocol level) & Digest-bound auth & Session escrow & N/A & Receipts + logs \\
MCP~\cite{research2024Model} & Tool sandboxing & Permission model & Server auth & N/A & Tool call logs \\
\bottomrule
\end{tabular}%
}
\end{table*}

No single protocol or execution interface covers all five dimensions. Payment and commerce protocols such as ERC-8004, AP2, OKX APP, x402, MPP, ACP, and ERC-8183 improve authorization, settlement, or inter-agent coordination, while MCP contributes tool-access control and auditability as an execution interface rather than a payment protocol. These mechanisms therefore remain complementary rather than sufficient: none of them by itself addresses LLM-layer compromise, long-horizon market manipulation, and regulatory compliance simultaneously. Framework-level agent systems are discussed elsewhere in the paper but are not co-scored here because they sit at a different abstraction layer from the protocols and interfaces compared here.

The marketplace side is similarly incomplete: emerging agent marketplaces and commerce layers promise discovery and settlement, but they still inherit unresolved problems around evaluator governance, listing integrity, and dispute resolution that earlier e-commerce work already warned about in human-mediated settings~\cite{rosen2025From, guttman1998Agent, nwana1998Agent}. Finance amplifies these weaknesses because rankings, escrow release, and reputation can all be monetized directly.

\subsection{Layered Defense Architecture}
\label{sec:trends}

The core design implication of our comparison is defense in depth across the full execution path:

\textbf{Layer 1: Prompt and Tool Hygiene.} Sanitize external inputs, tag agent-originated content, and verify tool provenance before invocation so public data and registry metadata cannot silently steer action selection~\cite{greshake2023Indirect, security2025Model}.

\textbf{Layer 2: Verified Execution Context.} Use output validation, runtime context binding, and capability graphs so that a locally plausible plan still has to match the current workflow, counterparty, and permitted action sequence before execution~\cite{ztrv2026Runtime, agentsentry2026Runtime}.

\textbf{Layer 3: Payment Authorization and Custody.} Separate cognition from custody, enforce scoped spending policies at the signing or credential layer, and bind payment or transaction parameters end to end using mechanisms such as ERC-8004 limits, AP2-style intents, OKX APP session keys and SDK-mediated payments, and x402/MPP request binding~\cite{goenka2026Tesspay, liu2026Agenticpay, okx2026AgentPaymentsProtocol, mpp2026Protocol}.

\textbf{Layer 4: Inter-Agent Trust Controls.} Require authenticated agent identity, stake- or reputation-backed evaluator selection, and anomaly monitoring for collusion or Sybil behavior in escrow-mediated commerce~\cite{virtuals2026ACP, virtuals2026ERC8183}.

\textbf{Layer 5: Market and Compliance Monitoring.} Add circuit breakers, cumulative position-drift detection, exposure aggregation, and tamper-evident audit trails so slow-burn manipulation and compliance abuse are visible before losses compound~\cite{faysal2026Agentic, dagostino2024Capturing}.

\subsection{Open Problems and Research Agenda}
\label{sec:gaps}
\label{sec:future}

Four research priorities follow from this condensed analysis. First, financial-agent benchmarks remain weak: existing agent-security and financial-LLM testbeds do not yet jointly capture irreversible execution, inter-agent settlement, and cumulative manipulation, which is why finance-specific benchmark harnesses are still needed~\cite{xie2024Finllm, son2024Krx, finvault2026Bench}. Second, long-horizon monitoring remains immature: O2P, N2C, and correlated-agent failures are cumulative and require metrics that work over days or weeks rather than per transaction~\cite{mony2025Llm, rizinski2026AiAgents}. Third, inter-agent trust still lacks a mature governance layer, especially around evaluator selection, cryptographic identity, and anti-collusion enforcement. Fourth, traditional payment rails and cross-chain deployments remain underexplored even though they introduce different reversibility, compliance, and custody assumptions from purely on-chain systems~\cite{hornuf2025Making}.

More specifically, general agent benchmarks such as AgentDojo and ASB still omit the finance-specific attack classes emphasized here, while newer financial testbeds such as TraderBench and CAIA improve adversarial realism but still do not fully integrate inter-agent trust and compliance vectors~\cite{agentdojo2024Benchmark, agentsecbench2025ASB, traderbench2026TraderBench, caia2025CAIA}. Financial LLM benchmarks in other languages and markets also broaden evaluation coverage but remain focused on capability rather than adversarial security~\cite{hirano2024Japanese, xu2024Superclue, lee2024Finance}. Theoretical work on verifiable reasoning, red-teaming, and longitudinal monitoring therefore remains directly relevant to this agenda~\cite{li2025Can, guo2024University, dong2024ScopingReview}.

Finally, some deployment trade-offs remain structurally hard rather than merely under-engineered. Portfolio agents can often tolerate stricter authorization and slower review loops than spot traders~\cite{ramirez2023ForAsset}, and adaptive policy tuning may partially reconcile autonomy with control~\cite{norman2004Adaptive}; but these mitigations do not remove the need for hard custody boundaries and market-level monitoring.

\section{Conclusion}
\label{sec:conclusion}

This paper provides a systematic account of the security challenges facing fully autonomous LLM agents in financial settings. As agentic finance matures through frameworks such as OpenClaw, payment and coordination protocols such as ERC-8004, AP2, OKX APP, x402, ACP, and MPP, and the broader convergence of LLMs with decentralized finance, these challenges will become more consequential rather than less.

The central result of this SoK is that autonomous financial-agent security is fundamentally a \emph{cross-layer} problem. Threats often originate in reasoning, tools, identity, or inter-agent interaction, but the resulting harm appears in custody, settlement, markets, or compliance. Our five-dimensional taxonomy provides a structured way to analyze this space, and the corpus synthesis shows that point defenses are not enough: secure deployment requires coordinated controls across agent integrity, authorization, trust, market structure, and regulation.

Several conclusions follow. First, agentic-finance risk is not simply the sum of traditional financial security and generic LLM security; it arises from their interaction under conditions of financial irreversibility and reduced human oversight~\cite{nie2024Survey, greshake2023Indirect}. Second, no existing system or protocol currently offers end-to-end coverage of this threat surface, which motivates the layered defense architecture developed in this paper. Third, common assumptions such as ``a human can intervene,'' ``prompt injection is a localized bug,'' or ``on-chain finality implies correctness'' do not hold for autonomous agents. Fourth, systemic risk from model and protocol homogeneity remains underappreciated: when many agents share the same foundation model or execution stack, a single exploit can propagate into market-wide disruption~\cite{moreno2025Predicting}.

The protocols being designed now will shape the infrastructure through which autonomous agents transact real value. Securing that infrastructure is a prerequisite for responsible deployment of autonomous AI in finance. We hope that the framework, corpus, attack taxonomy, defense architecture, and research agenda developed here provide a useful foundation for that effort.


\end{document}